\documentclass[revtex4]{emulateapj}
\usepackage{color}
\usepackage{courier}
\usepackage{sidecap}
\usepackage{natbib} 
\usepackage{longtable}
\usepackage{booktabs}
\usepackage{changepage}
\usepackage{amsmath}

\def\simlt{\lower.5ex\hbox{$\; \buildrel < \over \sim \;$}}
\def\simgt{\lower.5ex\hbox{$\; \buildrel > \over \sim \;$}}

\def\gsim{\lower 2pt \hbox{$\, \buildrel {\scriptstyle >}\over
{\scriptstyle \sim}\,$}}
\def\lsim{\lower 2pt \hbox{$\, \buildrel {\scriptstyle <}\over
{\scriptstyle \sim}\,$}}

\def\deg{\ifmmode ^{\circ}
         \else $^{\circ}$\fi}
\def\pdeg{\ifmmode
           $\setbox0=\hbox{$^{\circ}$}\rlap{\hskip.11\wd0 .}$^{\circ}
     \else \setbox0=\hbox{$^{\circ}$}\rlap{\hskip.11\wd0 .}$^{\circ}$\fi}
     
\def\pc{\ifmmode \mathrm{pc} \else $\mathrm{pc}$ \fi}
\def\mpc{\ifmmode \mathrm{Mpc} \else $\mathrm{Mpc}$\fi}
\def\mpcthree{\ifmmode \mathrm{Mpc}^{-3} \else $\mathrm{Mpc}^{-3}$\fi}
\def\gpcthree{\ifmmode \mathrm{Gpc}^{-3} \else $\mathrm{Gpc}^{-3}$\fi}

\def\kelvin{\ifmmode \mathrm{K} \else {$\mathrm{K}$}\fi}
\def\kev{\ifmmode \mathrm{keV} \else $\mathrm{keV}$ \fi}

\def\lsun{\ifmmode {L_\odot} \else $L_\odot$\fi}
\def\msun{\ifmmode M_\odot \else $M_\odot$\fi}
\def\msunyr{\ifmmode M_\odot~\mathrm{yr}^{-1} \else $M_\odot~\mathrm{yr}^{-1}$\fi}

\def\cosi{\ifmmode {\cos\,i} \else $\cos\,i$\fi}

\def\heii{\ifmmode {\rm He{\sc ii}} \else He~{\sc ii}\fi}
\def\mgii{\ifmmode {\rm Mg{\sc ii}} \else Mg~{\sc ii}\fi}
\def\caii{\ifmmode {\rm Ca{\sc ii}} \else Ca~{\sc ii}\fi}
\def\ciii{\ifmmode {\rm C{\sc iii}]} \else C~{\sc iii}]\fi}
\def\civ{\ifmmode {\rm C{\sc iv}} \else C~{\sc iv}\fi}
\def\mgii{\ifmmode {\rm Mg{\sc ii}} \else Mg~{\sc ii}\fi}

\newcommand{\oiii}{{\sc [O~iii]}}

\newcommand{\feii}{[Fe~{\sc ii}]}

\def\teff{\ifmmode {T_{\rm eff}} \else $T_{\rm eff}$\fi}
\def\tmax{\ifmmode {T_{\rm max}} \else $T_{\rm max}$\fi}

\def\mbh{\ifmmode {M_{\rm BH}} \else $M_{\rm BH}$\fi}
\def\led{\ifmmode L_{\mathrm{Ed}} \else $L_{\mathrm{Ed}}$\fi}
\def\lbolflare{\ifmmode L_{\mathrm{bol,flare}} \else $L_{\mathrm{bol,flare}}$\fi}
\def\lagn{\ifmmode L_{\mathrm{agn}} \else $L_{\mathrm{agn}}$\fi}
\def\lbolagn{\ifmmode L_{\mathrm{bol,agn}} \else $L_{\mathrm{bol,agn}}$\fi}
\def\lbol{\ifmmode L_{\mathrm{bol}} \else $L_{\mathrm{bol}}$\fi}
\def\mdot{\ifmmode {\dot M} \else $\dot M$\fi}
\def\mdoto{\ifmmode {\dot{M}_0} \else  $\dot{M}_0$\fi}
\def\mdotf{\ifmmode {\dot{M}_\mathrm{flare}} \else  $\dot{M}_\mathrm{flare}$\fi}

\def\hnot{\ifmmode H_0 \else H$_0$ \fi}

\def\vkep{\ifmmode v_\mathrm{Kep} \else $v_\mathrm{Kep}$ \fi}
\def\vc{\ifmmode v_\mathrm{c} \else $v_\mathrm{c}$ \fi}

\def\vthree{\ifmmode v_{1000} \else $v_{1000}$ \fi}
\def\vrel{\ifmmode v_\mathrm{rel} \else $v_\mathrm{rel}$ \fi}
\def\vkick{\ifmmode v_\mathrm{kick} \else $v_\mathrm{kick}$ \fi}
\def\vkickz{\ifmmode v_{\mathrm{kick},z} \else $v_{\mathrm{kick},z} $ \fi}
\def\vkicky{\ifmmode v_{\mathrm{kick},y} \else $v_{\mathrm{kick},y} $ \fi}
\def\vchar{\ifmmode v_\mathrm{char} \else $v_\mathrm{char}$ \fi}
\def\eflare{\ifmmode E_\mathrm{flare} \else $E_\mathrm{flare}$ \fi}
\def\ekick{\ifmmode E_\mathrm{kick} \else $E_\mathrm{kick}$ \fi}
\def\ecoll{\ifmmode E_\mathrm{coll} \else $E_\mathrm{coll}$ \fi}
\def\ezero{\ifmmode E_\mathrm{0} \else $E_\mathrm{0}$ \fi}
\def\efac{\ifmmode \xi_\mathrm{E} \else $\xi_\mathrm{E}$ \fi}
\def\tqso{\ifmmode t_\mathrm{QSO} \else $t_\mathrm{QSO}$ \fi}
\def\tflare{\ifmmode t_\mathrm{flare} \else $t_\mathrm{flare}$ \fi}
\def\tzero{\ifmmode t_\mathrm{0} \else $t_\mathrm{0}$ \fi}
\def\tfac{\ifmmode \xi_\mathrm{t} \else $\xi_\mathrm{t}$ \fi}
\def\gfac{\ifmmode f_\mathrm{g} \else $f_\mathrm{g}$ \fi}
\def\lflare{\ifmmode L_\mathrm{flare} \else $L_\mathrm{flare}$ \fi}
\def\fflare{\ifmmode F_\mathrm{flare} \else $F_\mathrm{flare}$ \fi}
\def\nflare{\ifmmode N_\mathrm{flare} \else $N_\mathrm{flare}$ \fi}
\def\tshock{\ifmmode T_\mathrm{shock} \else $T_\mathrm{shock}$ \fi}
\def\rmin{\ifmmode R_\mathrm{1} \else $R_\mathrm{1}$ \fi}
\def\rmax{\ifmmode R_\mathrm{2} \else $R_\mathrm{2}$ \fi}
\def\rbound{\ifmmode R_\mathrm{b} \else $R_\mathrm{b}$ \fi}
\def\pbound{\ifmmode P_\mathrm{b} \else $P_\mathrm{b}$ \fi}
\def\mbound{\ifmmode M_\mathrm{b} \else $M_\mathrm{b}$ \fi}
\def\mbo{\ifmmode M_{\mathrm{b}0} \else $M_{\mathrm{b}0} $ \fi}
\def\ebo{\ifmmode E_{\mathrm{b}0} \else $E_{\mathrm{b}0} $ \fi}
\def\efinal{\ifmmode E_\mathrm{final} \else $E_\mathrm{final} $ \fi}
\def\tbound{\ifmmode t_\mathrm{b} \else $t_\mathrm{b}$ \fi}
\def\tagn{\ifmmode t_\mathrm{AGN} \else $t_\mathrm{AGN}$ \fi}
\def\torb{\ifmmode t_\mathrm{orb} \else $t_\mathrm{orb}$ \fi}
\def\tdf{\ifmmode t_\mathrm{df} \else $t_\mathrm{df}$ \fi}
\def\rlim{\ifmmode R_\mathrm{lim} \else $R_\mathrm{lim}$ \fi}
\def\vlim{\ifmmode v_\mathrm{lim} \else $v_\mathrm{lim}$ \fi}
\def\vphi{\ifmmode v_\phi \else $v_\phi$ \fi}
\def\mlim{\ifmmode M_\mathrm{lim} \else $M_\mathrm{lim}$ \fi}
\def\tlim{\ifmmode t_\mathrm{lim} \else $t_\mathrm{lim}$ \fi}
\def\llim{\ifmmode L_\mathrm{lim} \else $L_\mathrm{lim}$ \fi}
\def\fqso{\ifmmode f_\mathrm{QSO} \else $f_\mathrm{QSO}$ \fi}

\def\hbeta{\ifmmode \rm{H}\beta \else H$\beta$\fi}
\def\hbetan{\ifmmode \rm{H}\beta_{\rm n} \else H$\beta_{\rm n}$\fi}
\def\hgamma{\ifmmode \rm{H}\gamma \else H$\gamma$\fi}
\def\hdelta{\ifmmode \rm{H}\delta \else H$\delta$\fi}
\def\hepsilon{\ifmmode \rm{H}\epsilon \else H$\epsilon$\fi}
\def\hzeta{\ifmmode \rm{H}\zeta \else H$\zeta$\fi}
\def\halpha{\ifmmode \rm{H}\alpha \else H$\alpha$\fi}
\def\lalpha{\ifmmode \rm{Ly}\alpha \else Ly$\alpha$}

\def\dvhb{\ifmmode \Delta v_{\hbeta} \else $\Delta v_{\hbeta}$\fi}
\def\dvmg{\ifmmode \Delta v_{\rm{Mg}} \else $\Delta v_{\rm{Mg}}$\fi}

\def\muobs{\ifmmode {\mu_{o}} \else  $\mu_{o}$ \fi}
\def\cosi{\ifmmode {\mathrm{cos}\,i} \else $\mathrm{cos}\,i$\fi}

\def\teff{\ifmmode {T_{eff}} \else $T_{eff}$ \fi}
\def\tmax{\ifmmode {T_{max}} \else $T_{max}$ \fi}

\def\tauh{\ifmmode {\tau_{\rm H}} \else $\tau_{\rm H}$ \fi}

\def\yr{\ifmmode {\rm yr} \else  yr \fi}
\def\kms{\ifmmode \rm km~s^{-1}\else $\rm km~s^{-1}$\fi}
\def\cm{\ifmmode {\rm cm} \else  cm \fi}
\def\cmmitwo{\ifmmode \rm cm^{-2} \else $\rm cm^{-2}$\fi}
\def\cmmithree{\ifmmode \rm cm^{-3} \else $\rm cm^{-3}$\fi}
\def\cmps{\ifmmode \rm cm~s^{-1}\else $\rm cm~s^{-1}$\fi}
\def\cmpsps{\ifmmode \rm cm~s^{-2}\else $\rm cm~s^{-2}$\fi}
\def\kmps{\ifmmode \rm km~s^{-1}\else $\rm km~s^{-1}$\fi}
\def\kmpspmpc{\ifmmode \rm km~s^{-1}~Mpc^{-1} \else
    $\rm km~s^{-1}~Mpc^{-1}$\fi}
  
\def\gcmthree{\ifmmode \rm g~cm^{-3} \else $\rm g~cm^{-3}$\fi}
\def\gcmtwo{\ifmmode \rm g~cm^{-2} \else $\rm g~cm^{-2}$\fi}
   
\def\erg{\ifmmode {\rm erg} \else $\rm erg$ \fi}
\def\ergps{\ifmmode {\rm erg~s^{-1}} \else $\rm erg~s^{-1}$ \fi}
\def\ergcms{\ifmmode \rm erg~cm^{-2}~s^{-1} \else $\rm erg~cm^{-2}~s^{-1}$ \fi}
\def\ergcmshz{\ifmmode \rm erg~s^{-1}~cm^{-2}~Hz^{-1} \else $\rm
erg~cm^{-2}~s^{-1}~Hz^{-1}$ \fi}
\def\ergcmsa{\ifmmode \rm erg~cm^{-2}~s^{-1}~\AA^{-1} \else $\rm
erg~cm^{-2}~s^{-1}~\AA^{-1}$ \fi}
\def\ergshz{\ifmmode \rm erg s^{-1} Hz^{-1} \else
   $\rm erg s^{-1} Hz^{-1}$ \fi}

\def\lam{\ifmmode {\lambda} \else {$\lambda$} \fi}
\def\llam{\ifmmode {L_\lambda} \else  $L_\lambda$ \fi}
\def\lamLlam{\ifmmode \lambda L_{\lambda}(5100) \else {$\lambda L_{\lambda}(5100)$} \fi}
\def\nuLnu{\ifmmode \nu L_{\nu}(5100) \else {$\nu L_{\nu}(5100)$} \fi}
\def\ilam{\ifmmode {I_\lambda} \else  $I_\lambda$ \fi}
\def\flam{\ifmmode {F_\lambda} \else  $F_\lambda$ \fi}
\def\inu{\ifmmode {I_\nu} \else  $I_\nu$ \fi}
\def\fnu{\ifmmode {F_\nu} \else  $F_\nu$ \fi}
\def\bnu{\ifmmode {B_\nu} \else  $B_\nu$ \fi}

\def\msigma{\ifmmode M_{\sigma} \else $M_{\sigma}$\fi}
\def\mbulge{\ifmmode M_{\mathrm{bulge}} \else $M_{\mathrm{bulge}}$\fi}
\def\mgal{\ifmmode M_{\mathrm{gal}} \else $M_{\mathrm{gal}}$\fi}
\def\lgal{\ifmmode L_{\mathrm{gal}} \else $L_{\mathrm{gal}}$\fi}
\def\lbulge{\ifmmode L_{\mathrm{bulge}} \else $L_{\mathrm{bulge}}$\fi}
\def\mgalstar{\ifmmode M^*_{\mathrm{gal}} \else $M^*_{\mathrm{gal}}$\fi}

\def\mbhsigstar{\ifmmode M_{\mathrm{BH}} - \sigma_* \else $M_{\mathrm{BH}} - \sigma_*$ \fi}
\def\deltalogmbh{\ifmmode \Delta~{\mathrm{log}}~M_{\mathrm{BH}} \else $\Delta$~log~$M_{\mathrm{BH}}$\fi}

\def\sigstar{\ifmmode \sigma_* \else $\sigma_*$\fi}
\def\sigthree{\ifmmode \sigma_{\mathrm{[O~III]}} \else $\sigma_{\mathrm{[O~III]}}$\fi}
\def\sigtwo{\ifmmode \sigma_{\mathrm{[O~II]}} \else $\sigma_{\mathrm{[O~II]}}$\fi}
\def\signl{\ifmmode \sigma_{\mathrm{NL}} \else $\sigma_{\mathrm{NL}}$\fi}
\def\wthree{\ifmmode {\rm FWHM({[O~III]})} \else $FWHM({[O~III]})$ \fi}
\def\wtwo{\ifmmode {\rm FWHM({[O~II]})} \else $FWHM({[O~II]})$ \fi}
\def\mthree{\ifmmode M_{\mathrm [O~III]} \else $M_{\mathrm [O~III]}$ \fi}
\def\mtwo{\ifmmode M_{\mathrm [O II]} \else $M_{\mathrm [O II]}$ \fi}
\def\lbreak{\ifmmode L_{\mathrm{break}} \else $L_{\mathrm{break}}$\fi}
\def\lcut{\ifmmode L_{\mathrm{cut}} \else $L_{\mathrm{cut}}$\fi}

\shortauthors{Smith et al.}
\shorttitle{\emph{Kepler} Timing of AGN}

\begin{document}

\title{Evidence for an Optical Low-frequency Quasi-Periodic Oscillation in the \emph{Kepler} Light Curve of an Active Galaxy}

\author{Krista Lynne Smith\altaffilmark{1,2}, Richard~F.~Mushotzky\altaffilmark{2}, Patricia~T.~Boyd\altaffilmark{3} \& Robert V. Wagoner\altaffilmark{4}}

\altaffiltext{1}{Einstein Fellow, KIPAC at SLAC, Stanford University; klynne@stanford.edu}

\altaffiltext{2}{University of Maryland, College Park, MD 20742}

\altaffiltext{3}{NASA/GSFC, Greenbelt, MD 20771, USA}

\altaffiltext{4}{Department of Physics and KIPAC, Stanford University}

\begin{abstract}

We report evidence for a quasi-periodic oscillation (QPO) in the optical light curve of KIC~9650712, a narrow-line Seyfert~1 galaxy in the original \emph{Kepler} field. After the development and application of a pipeline for \emph{Kepler} data specific to active galactic nuclei (AGN), one of our sample of 21 AGN selected by infrared photometry and X-ray flux demonstrates a peak in the power spectrum at log~$\nu=-6.58$~Hz, corresponding to a temporal period of $t=44$~days. We note that although the power spectrum is well-fit by a model consisting of a Lorentzian and a single power law, alternative continuum models cannot be ruled out. From optical spectroscopy, we measure the black hole mass of this AGN as log~($M_{\mathrm{BH}}/M_\odot) = 8.17$. We find that this frequency lies along a correlation between low-frequency QPOs and black hole mass from stellar and intermediate mass black holes to AGN, similar to the known correlation in high-frequency QPOs. 

\end{abstract}

\section{Introduction}
\label{sec:intro}
Quasi-periodic oscillations (QPOs) have been seen in the X-ray power spectra of the majority of stellar mass black hole candidates in X-ray binaries \citep{Remillard2006}. These oscillations belong to two main types: low- and high-frequency QPOs. High-frequency QPOs are the rarer type. They occur in the range of tens to hundreds of Hz, and have often been found to manifest in pairs with a 3:2 frequency ratio. Low-frequency QPOs are stronger and more ubiquitous than high-frequency QPOs. They occur in the frequency range of mHz to $\sim30$~Hz, and can drift in centroid frequency. More details on these properties can be found in the reviews by \citet{Remillard2006} and \citet{Motta2016}. 

There are many physical origin theories for QPOs. The behavior underlying these phenomena is believed to occur very near to the black hole itself, perhaps within a few gravitational radii. High-frequency QPOs have been proposed as consequences of periastron and orbital disk precession \citep{Stella1999}, warped accretion disks \citep{Kato2005}, global disk oscillations \citep{Titarchuk2000}, magnetic reconnection \citep{Huang2013}, magnetically-choked accretion flows \citep{McKinney2012} and diskoseismology \citep{Wagoner2001}. The origin of low-frequency QPOs varies depending on their detailed type \citep[A, B, or C; see ][]{Motta2016}, and include unstable spiral density waves \citep{Tagger1999}, viscous magneto-acoustic oscillations in a spherical transition layer near the compact object \citep{Titarchuk2004}, and Lense-Thirring precession \citep[e.g.,][]{Ingram2009}. Regardless of the mechanism responsible for these rapid variations, their origin in accretion-related structures very near the black hole makes them a rare and valuable probe of strong gravity and the effect of black holes on their immediate environments. 

The first QPO in an AGN was discovered in the X-ray light curve of RE~J1034+396 by \citet{Gierlinski2008} and robustly confirmed by \citet{Alston2014}. Recently, X-ray QPOs have been detected in two intermediate-mass black hole (IMBH) candidates and a handful of additional active galactic nuclei (AGN). A remarkable linear correlation of the frequency of the QPO and the black hole mass seems to hold over many orders of magnitude, from stellar mass black holes with $M\sim10~M_{\odot}$, to supermassive black holes of $M\sim10^6~M_{\odot}$~\citep{Abramowicz2004, Zhou2015}. The universality of this correlation links accretion processes across vast scales, and indicates that QPO frequency may act as a very accurate probe of black hole mass. As previous authors have indicated, such a $1/M$~scaling is indeed expected if the oscillations are in any way dependent on the characteristic length scale of strong gravity. Interestingly, all of the AGN QPO candidates are in a spectroscopic subclass known as Narrow-Line Seyfert 1s (NLS1). These objects are characterized by relatively narrow broad emission lines (FWHM$\leq2000~\mathrm{km~s}^{-1}$), strong Fe~II emission, and weak [O~III] emission compared to H$\beta$. Such objects may have very high accretion rates; see the review by \citet{Komossa2007}.


\begin{figure*}
    \centering
    \includegraphics[width=\textwidth]{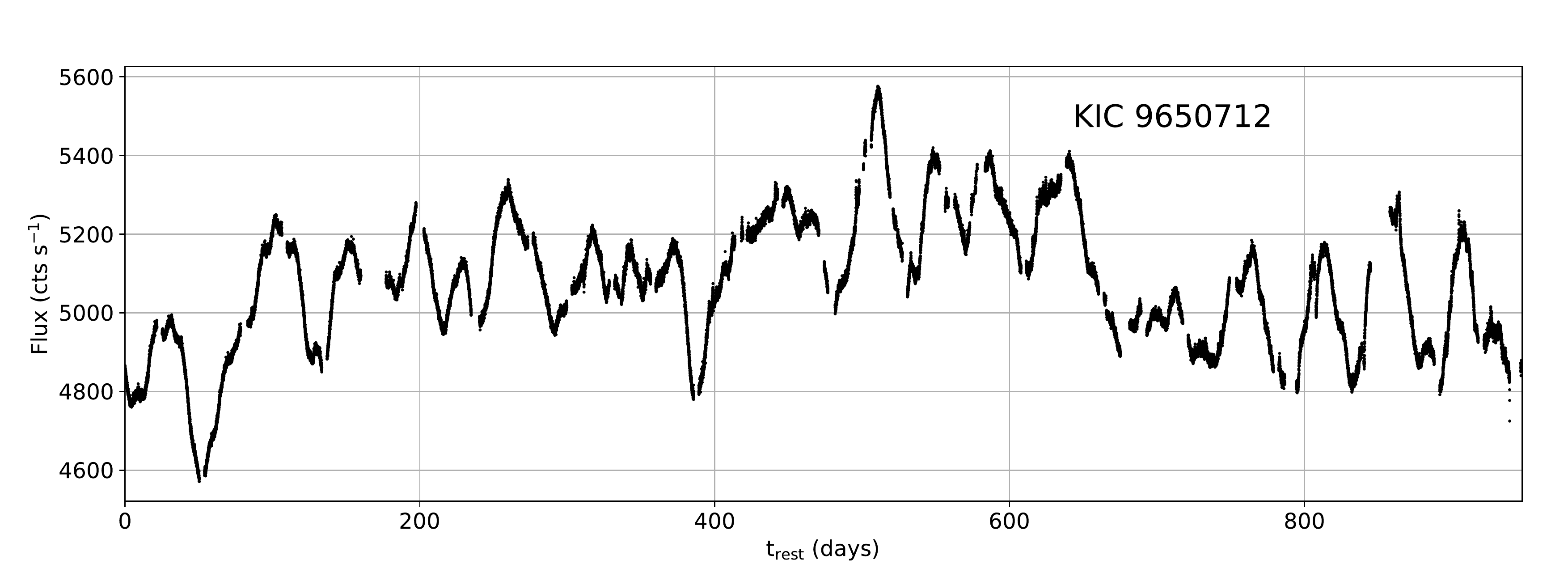}
    \caption{\emph{Kepler} light curve of KIC~9650712 after processing by a specialized AGN pipeline. Each 30-minute cadence data point is shown. The typical flux error on each data point is approximately 4 counts. For information on gaps and reduction, see \citet{Smith2018}.}
    \label{fig:qpo_lc}
\end{figure*}


So far, no optical QPO has been reported in an AGN. This is perhaps because ground-based, sporadically sampled optical light curves have never been directly comparable to the high-precision X-ray light curves generated by space telescopes. Fortunately, the unparalleled precision and regular sampling of the \emph{Kepler} exoplanet satellite has lately produced precise, evenly-sampled space-based optical light curves. Our team has developed a sophisticated pipeline for the treatment of \emph{Kepler} AGN light curves \citep{Smith2018}. Among the phenomena revealed by our approach is the first optical QPO candidate in an AGN.  Searching for periodicities with the sparse and uneven sampling of ground based telescopes is problematic, since the red noise nature of AGN variability can easily mimic periodic signals \citep{Vaughan2016}. The \emph{Kepler} light curves can be analyzed with Fourier techniques that enable period detection in the frequency domain. While more robust than time domain detection, there is still a risk of red noise mimicking our periodic signal; however, our candidate is detected as a peak in the Fourier power spectrum of an AGN matching the spectroscopic sub-type of all current X-ray AGN QPOs and agrees very well with the extrapolation of an existing correlation between QPO frequency and black hole mass.

\section{Power Spectrum Modeling}
\label{sec2}
This object is part of a sample of 21 Type 1 AGN monitored by the \emph{Kepler} spacecraft during its 3.5 year mission with 30-minute cadence, selected using a combination of infrared photometric techniques \citep{Edelson2012} and X-ray detection \citep{Smith2015}. While summarized here, the full details of this sample, the special methods necessary for analyzing \emph{Kepler} AGN data, and the reduction methods used can be found in \citet{Smith2018}. The resulting light curve for our QPO candidate, KIC~9650712, is shown in Figure~\ref{fig:qpo_lc}. We obtained an optical spectrum of this target from Lick Observatory, and calculated the redshift to be $z=0.128$. The light curve of KIC~9650712 spans 950 days in the object's rest frame. We have also used the FWHM of the H$\beta$ emission line to calculate the black hole mass, a method that is very commonly used for AGN and for QPO X-ray candidates in particular. Based on the accepted formulae from \citet{Wang2009}, we estimate a mass of log~($M_{\mathrm{BH}}/M_\odot) = 8.17\pm0.20$, two orders of magnitude larger than the most massive object in the small number of known AGN QPOs. We conservatively assume the larger error estimate on this method found by \citet{Vestergaard2006} of $\sim0.5$~dex. Although this mass is higher than the usual mass for the NLS1 class, the FWHM of H$\beta$~is only 2270~km~s$^{-1}$, much lower than the other Type 1 AGN in our sample. In order to calculate the Eddington ratio, we first estimate the bolometric luminosity using the \emph{Swift} survey value of $L_X = 1.62\times10^{44}~\mathrm{erg s}^{-1}$ and the hard X-ray bolometric correction of \citet{Vasudevan2007}. To be consistent with many of the other optical studies used here, we also perform the calculation using the correction on $L_{5100}$ from \citet{Runnoe2012}. The two measurements of $L / L_\mathrm{Edd}$ are 0.14 and 0.23, respectively. The latter estimate makes it the highest accretion rate object in our sample of \emph{Kepler} AGN. Additionally, the spectrum shows very strong \feii~emission, and we measure the \oiii/H$\beta$~line ratio to be 0.14, a value comfortably less than the definition threshold for NLS1s of \oiii/H$\beta < 3$. We mention this because all of the current X-ray QPO candidates happen to be in NLS1 galaxies.

\begin{figure}
\begin{tabular}{c}

\includegraphics[width=0.5\textwidth]{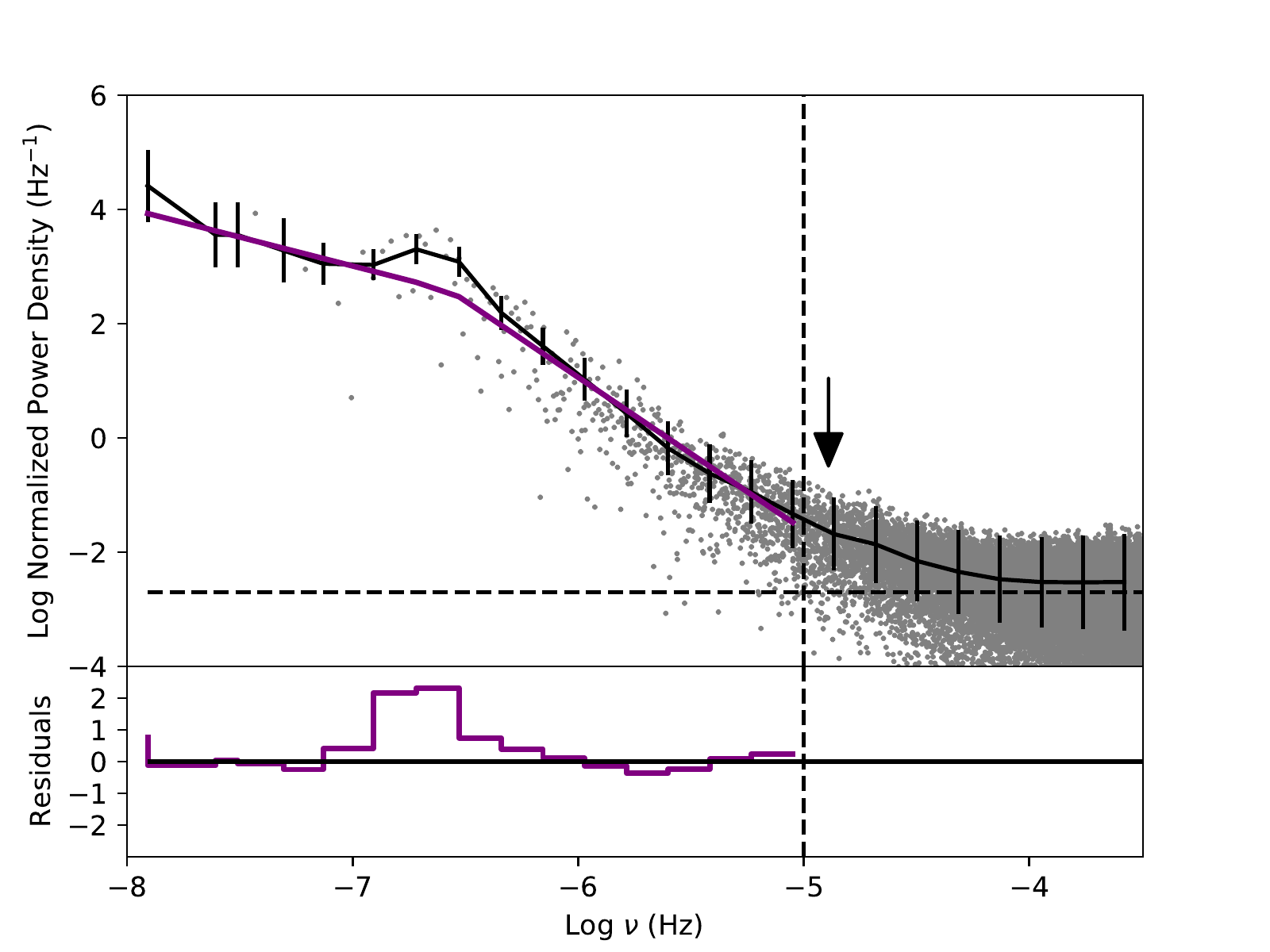} \\ \includegraphics[width=0.5\textwidth]{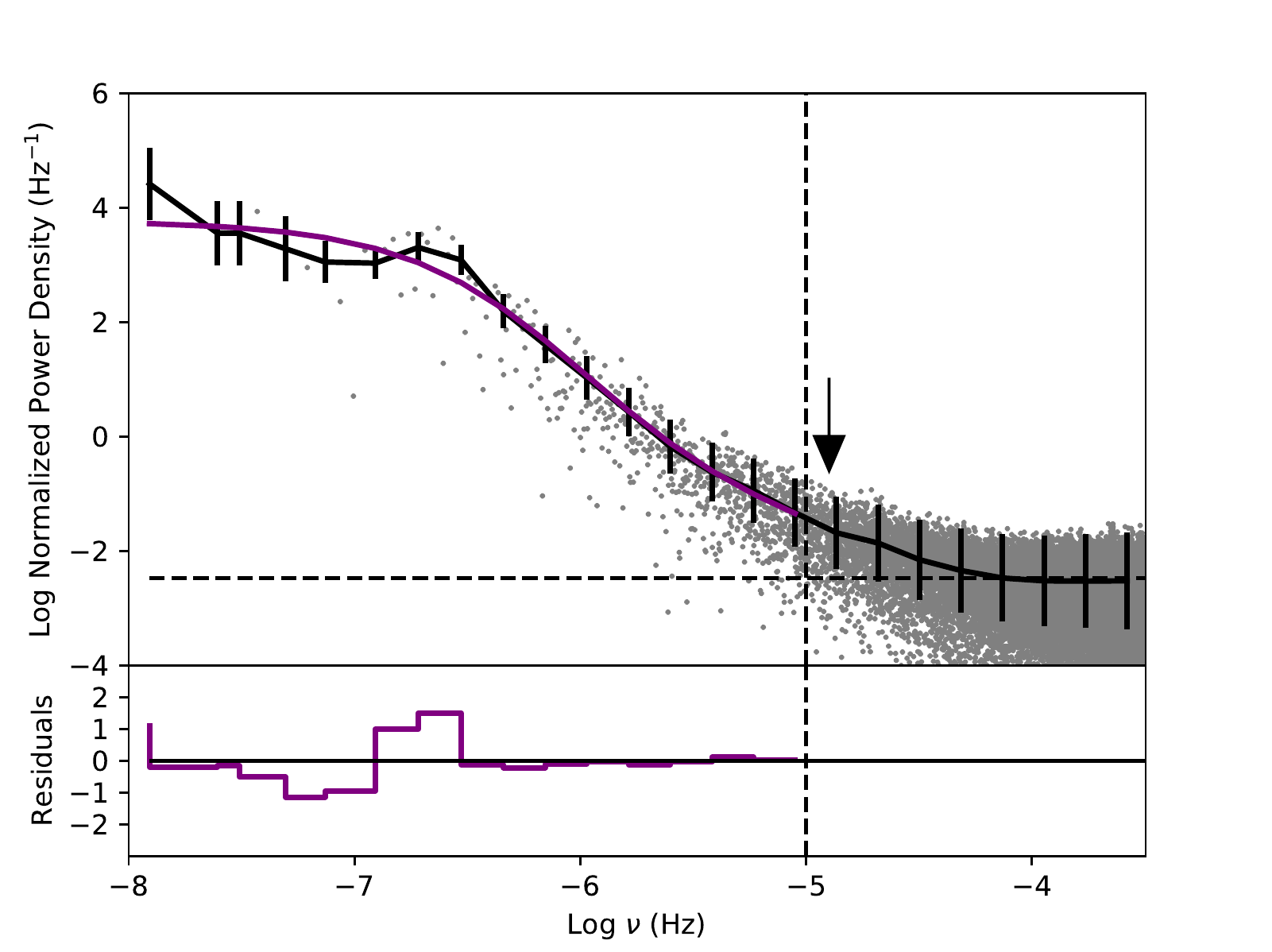} \\ \includegraphics[width=0.5\textwidth]{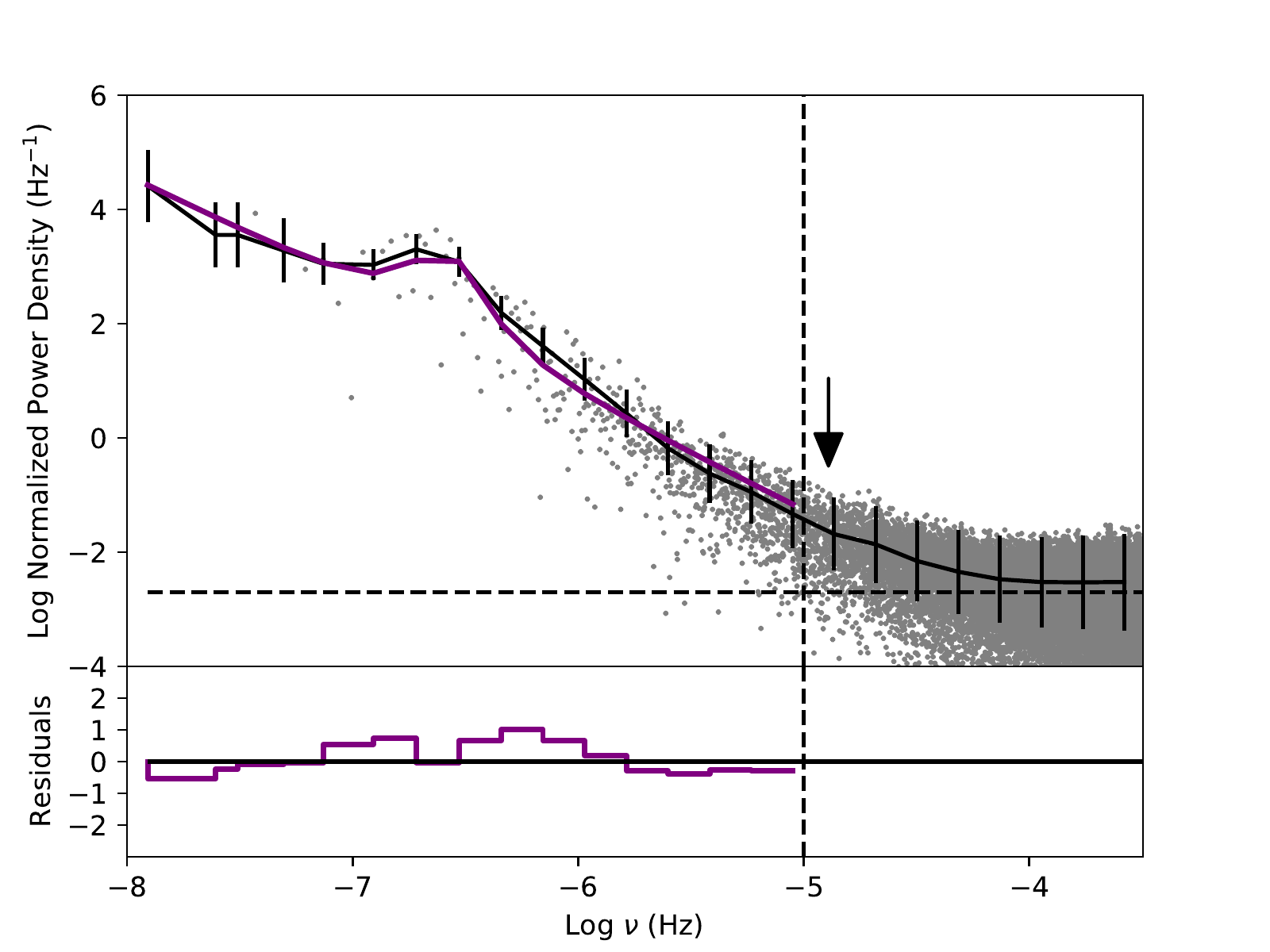} \\

\end{tabular}
\caption{Power spectrum of KIC~9650712 both raw (grey dots) and binned (solid black). Error bars correspond to the rms-spread of simulated light curves. The top panel shows the fit and residuals for a broken power law, the middle for a bending power law, and the bottom panel for a single power law plus Lorentzian. The horizontal dashed line represents the value for the expected Poisson noise; fitting is only performed for frequencies below this value. The arrows denote the location where the corresponding high-frequency QPO would be; see discussion at the end of Section~\ref{sec:corr}.}
\label{fig:qpo_powspec}
\end{figure}

Our aim in \citet{Smith2018} was to detect characteristic timescales in AGN and determine the power spectral slope at high frequencies. The red-noise power spectra of AGN are typically well-fit by a power law, where the spectral density $S$ varies with the frequency as $S \propto f^{-\alpha}$. In order to determine whether our objects were well-fit by a single power law, we followed the PSRESP process, described in \citet{Uttley2002}. Briefly, this consists of simulating a very long light curve from a given power spectral slope using the \citet{Timmer1995} algorithm, which allows 500 light curves of the observed length to be drawn from it without overlap. The same gaps and interpolation techniques are introduced and used on the simulated light curves as on the original, and 500 power spectra are created from the simulated light curves. The rms spread of these power spectra become the error bars on the observed power spectrum. The observed spectrum is then fit above the Poisson cutoff value by a single power law, generating a $\chi^2$~value. The goodness-of-fit is then measured by calculating the percentile value above which the observed $\chi^{2}$ exceeds the simulated distribution. This percentage is the confidence with which we can reject the model. In a few cases, single power-law models were always rejected with high confidence at all slopes ranging from $\alpha=1.5-3.5$. In the case of KIC~9650712, a single power law model with the observed slope can be rejected with 83\% confidence: an acceptable fit, but one that could perhaps be improved upon. In our initial analysis, a broken power law model provides an acceptable fit to the data with a high-frequency slope of $\alpha=2.9$. However, we found that this object experienced the highest $\chi^{2}$~minimization when fit with a single power law with a slope of $\alpha=1.9$ and a Lorentzian component.

In order to explore the goodness-of-fit of other possible models in more detail, we compare a broken power law, bending power law, and single power law plus a Lorentzian component to the single power law case. Our broken power law modeling procedure can be found in \citet{Smith2018}, our bending power law model corresponds to Equation~3 of \citet{Gonzalez2012}, and our periodic model consists of the sum of a linear component (the underlying single power law) and a standard Lorentzian. The power spectra and residuals for the broken power law, bending power law, and quasi-periodic Lorentzian are shown in Figure~\ref{fig:qpo_powspec}. We have normalized the power spectra by a constant $A_{\mathrm{rms}}^{2} = 2\Delta T_{\mathrm{samp}}/\bar{x}^{2}N$, where $\Delta T_{\mathrm{samp}}$ is the sampling interval, $\bar{x}$ is the mean count rate in cts~s$^{-1}$, and $N$ is the total number of data points \citep{VanderKlis1997}. 

\begin{figure}
    \centering
    \includegraphics[width=0.5\textwidth]{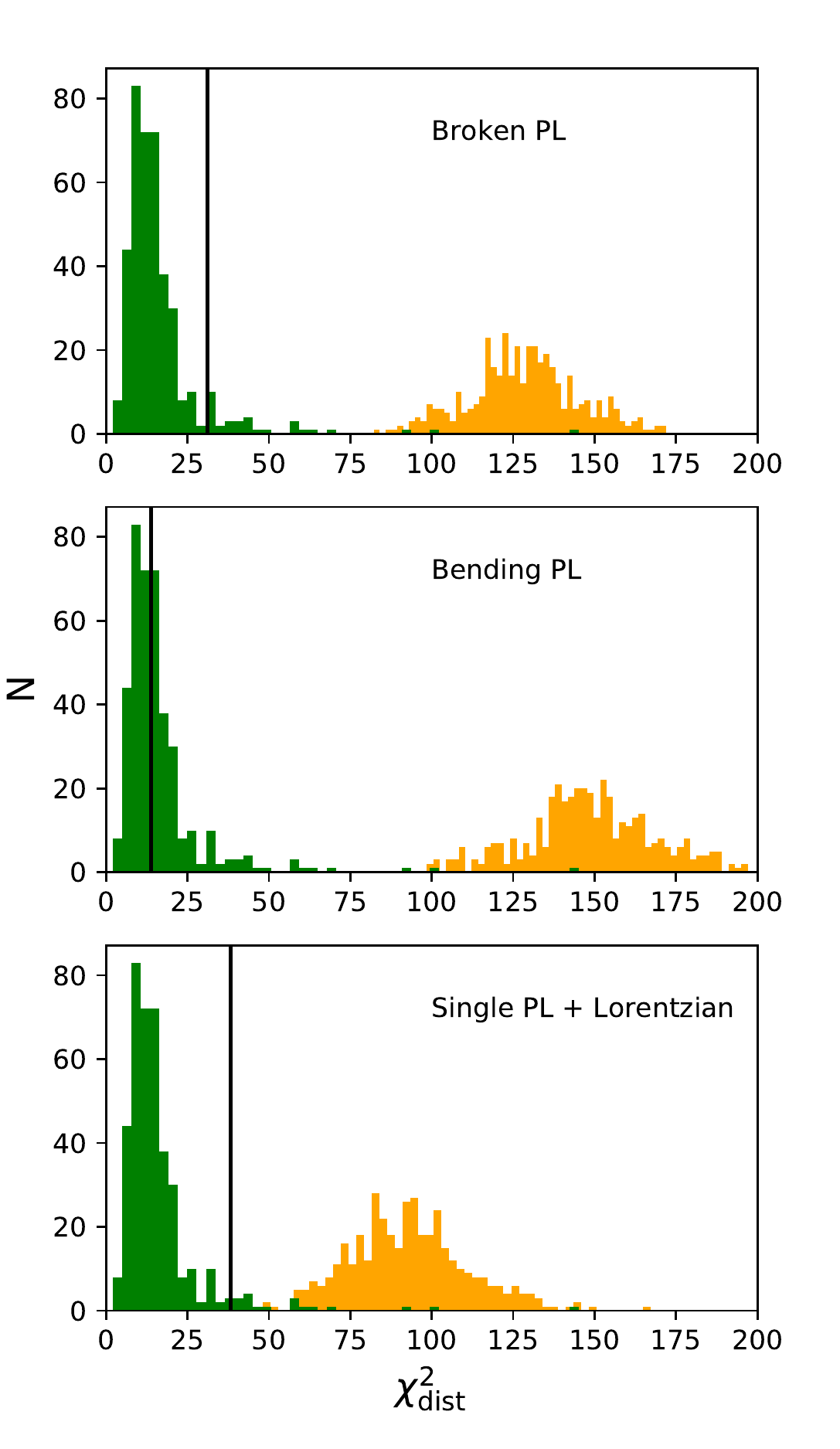}
    \caption{Distributions of the fiducial $\chi^{2}_{\mathrm{dist}}$ (green) and the $\chi^{2}_{\mathrm{dist}}$ values from comparing each simulated realization of a single power law to the broken, bending, and single + Lorentzian models (orange). The $\chi^{2}_{\mathrm{dist}}$ value of the observed data compared to the model is shown by a black line.}
    \label{fig:chisqs}
\end{figure}

To determine whether or not these models provide better fits than a single power law in reality or simply because they have more free parameters, we follow the method of \citet{Summons2007}. Using the 500 light curves simulated from the best-fitting single power law slope, we fit each simulated light curve with the best-fitting broken power law parameters, bending power law parameters, and single power law plus Lorentzian parameters, just as was done with the real data. We then compare histograms of these fit probabilities to the fiducial $\chi^2$ distribution calculated as described in \citet{Smith2018}. In Figure~\ref{fig:chisqs} we show the three comparisons. Note that the $\chi^{2}$ values given here differ slightly from those in \citet{Smith2018}; this is because each time the PSRESP process is run, a slightly different ensemble of simulated light curves is generated, resulting in slightly different error bars.  In all cases, no given simulation has a better fit to the more complex model than the observed power spectrum. Each of these models provides a better fit than a single power law model, but all are acceptable fits to the data. A single power law plus a periodic component is then just one of several complex models that provide good fits. As a consistency check on our periodic model, we have also computed the Lomb-Scargle periodogram on the processed light curve without including any linear interpolation, since this method is capable of handling unevenly sampled data \citep{Scargle1982}. The periodogram also shows a peak at $\sim45$ days.

The best-fitting Lorentzian model has a central frequency of log~$\nu=-6.58$~Hz, corresponding to a temporal period of $t=44$~days. The $Q$-value of coherence of the feature, defined as $\nu_0 / \mathrm{FWHM}$ \citep{Nowak1999}, is $Q=1.69$; somewhat lower than low-frequency QPOs in the literature. Possible reasons for this can be found in the conclusion. Papers reporting X-ray QPOs frequently calculate the fractional rms of the periodic component. In our case, such a measurement would be misleading, since the host galaxy contributes a large constant flux to the \emph{Kepler} bandpass that cannot be determined from the present data. The quoted X-ray values do not suffer from a contaminating constant term, and so could not be compared with our value.


\section{A Correlation of Black Hole Mass and Low-Frequency QPOs}
\label{sec:corr}

\begin{figure*}
    \centering
    \includegraphics[width=0.8\textwidth]{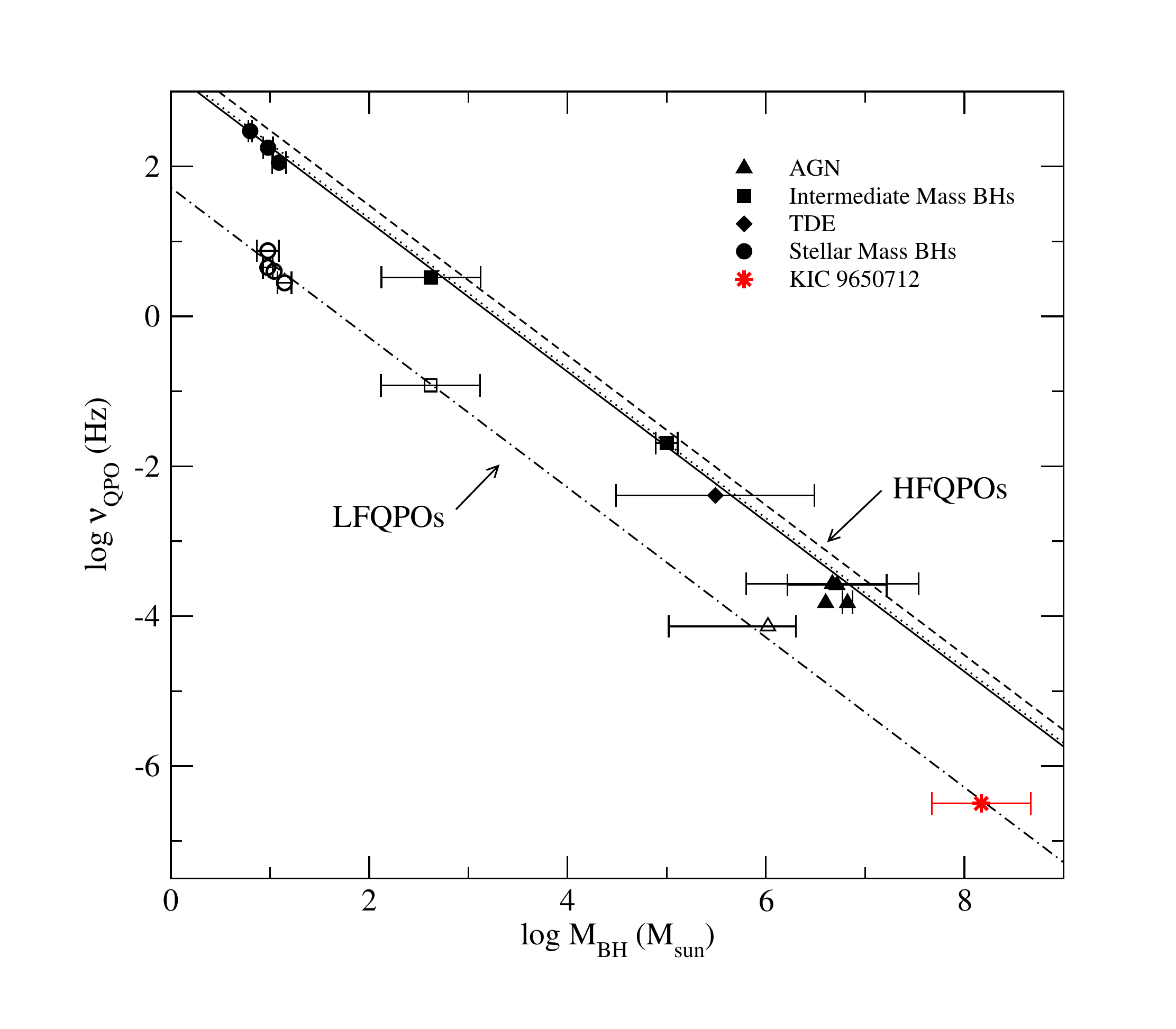}
    \vspace{-10mm}
    \caption{Known high- and low-frequency QPOs and their relationships with black hole mass. Solid and hollow symbols represent high- and low-frequency QPO points, respectively (the frequency is the stronger 2$\times \nu_0$ peak of the 3:2 pair). The solid line represents the relationship for stellar mass black holes from \citet{Remillard2006}. The dotted and dashed lines show the relationships derived from 3:2  and 3:1 resonance models by \citet{Aschenbach2004a}. The dot-dashed line is our linear regressive fit, preserving the $1/M$~slope, through only the stellar and intermediate mass points not including our candidate. The LFQPOs in stellar mass black holes are XTE~J1550-564 \citep{Vignarca2003}, GRS~1915+105 \citep{Vignarca2003,Strohmayer2009}, XTE~J1859+226 \citep{Filippenko2001}, and H~1743-322 \citep{McClintock2009}. The intermediate mass black holes are M~82 X-1, which exhibits both a low and high frequency QPO \citep{Dewangan2006,Pasham2014} and NGC~5408 X-1, which has only a HFQPO reported \citep{Strohmayer2009,Huang2013}. The NLS1 AGN in the figure are Mrk~766 \citep{Zhang2017}, RE~J1034+396 \citep{Gierlinski2008}, 1H707-495 \citep{Pan2016}, and MS~2254.9-3712 \citep{Alston2015}. The tidal disruption event (TDE) is Swift~J1644+57 \citep{Reis2012}. The lone LFQPO AGN detection is 2XMM~J123103.2+110648 \citep{Lin2013}.
    \vspace{9mm}}
    \label{fig:qpombh}
\end{figure*}

There is now considerable evidence that the correlation between black hole mass and the central frequency of the 2$\times \nu_0$~peak in high-frequency QPOs extends from stellar masses to supermassive black holes with $M_\mathrm{BH}\sim10^{6}~M_\odot$ \citep{Remillard2006,Abramowicz2004,Zhou2015}. In stellar and intermediate mass black holes, low-frequency QPOs can be used in tandem with spectral variations to predict the black hole mass \citep{Fiorito2004,Casella2008}. When both the high- and low-frequency QPOs have been detected in a given object, the two methods provide independent checks on the mass. This is the case for several stellar mass black holes and the intermediate mass ULX in M~82. In the case of AGN, independent mass checks are provided by other measurement methods. The most frequently used is the H$\beta$~width method employed here, but Mrk~766 has a very well-determined mass from reverberation mapping \citep{Bentz2010}. When the independent mass measurement agrees with the prediction from the high-frequency correlation, and especially if the observed QPO has the well-known 3:2 ratio of high-frequency QPOs, one can confidently claim the detected QPO is high-frequency. In Figure~\ref{fig:qpombh}, we reproduce the plot from \citet{Zhou2015} and \citet{Abramowicz2004} for the high-frequency QPOs, and add the known low-frequency QPOs. The references for the high-frequency points can be found in those papers, except for the more recent detections described in the figure caption. We note that MS~$2254.9-3712$ and Mrk~766 may exhibit QPOs in a 3:2 frequency ratio, strengthening their high-frequency classification. We note also that \citet{Alston2014} has pointed out that searching for transient QPO phenomena via data mining, as has been done for some of the AGN QPOs, is potentially problematic; however, we report them here for completeness. The lines in the plot correspond to the previously-observed relation and the resonance models of \citet{Aschenbach2004a}, which translate vertically depending on black hole spin.

The current roster of low-frequency QPOs with independent mass estimates includes the stellar mass black holes XTE~J1550-564 \citep{Vignarca2003}, GRS~1915+105 \citep{Vignarca2003}, XTE~J1859+226 \citep{Casella2004}, and H~1743-322 \citep{McClintock2009}, the intermediate mass black hole M82~X1 \citep{Strohmayer2003,Dewangan2006}, and the lone low-frequency QPO detection in an AGN, 2XMM~J123103.2+110648 \citep{Lin2013}. We note that the mass of this last object is uncertain. It was determined by \citep{Ho2012} to be very low for an AGN ($\sim10^{5} M_\odot$), using the M-$\sigma$ relation since the object is a Type~2 AGN (i.e., it does not exhibit Doppler-broadened emission lines, precluding the H$\beta$ method). The validity of the M-$\sigma$~relation is in doubt for AGN and for NLS1s in particular. The mass was later determined by \citet{Lin2013} to be $2\times10^{6} M_\odot$~based on X-ray and UV spectral fitting. For plotting, we use the average of these two estimates with the error bar encompassing the full range of both. 

The mass of M82~X-1, while previously quite uncertain, has now been calculated using both the high-frequency QPO extrapolation and a relativistic precession model with consistent results at $\sim420 M_\odot$ \citep{Pasham2014}. Nevertheless, we show a large error bar on the measurement to encompass the range of the mass estimated from the low-frequency QPO scaling with spectral index done by \citet{Casella2008}, $95-1260 M_\odot$, so that the reader may see the results of both methods. 

The case of the intermediate mass ULX NGC~5408~X-1 is not as well determined. A QPO has been detected robustly at $20$~mHz \citep{Strohmayer2009}, but whether it is a low- or high-frequency QPO is not known, and there is no independent mass measurement. \citet{Huang2013} quote the mass as $10^{5} M_\odot$ based on the assumption of the object as a high-frequency QPO and fitting of the X-ray spectrum, while the original analysis by \citet{Strohmayer2009} claim it is a low-frequency QPO, and use the previously mentioned \citet{Fiorito2004} method to calculate a mass of $2000-5000 M_\odot$. We have placed the object among the high-frequency QPOs in the plot because the \citet{Huang2013} method used X-ray spectral fitting to back up their mass estimate. It does not affect the veracity of the plot one way or the other, since the mass estimates arise only from the assumption of the QPO type and so would fall on either correlation by design. 

Given the mass of log~($M_{\mathrm{BH}}/M_\odot) = 8.17\pm0.20$ measured using the FWHM of the H$\beta$~line, the predicted mass is approximately a factor of 40 below the high-frequency QPO correlation. However, a factor of 40 is approximately the difference between the central frequencies of low- and high-frequency QPOs in X-ray binaries. Motivated by this coincidence, we plotted our candidate QPO with the low-frequency QPOs of X-ray binaries and intermediate mass black hole candidates. If we perform a linear regressive fit through the stellar and intermediate mass points only, preserving the $1/M$~dependence, we obtain:

\begin{equation}
f\mathrm{(Hz)} = 51.9~(M_\mathrm{BH} / M_\odot)^{-1}.
\end{equation}

If we extrapolate this all the way to our mass, we find that our object is in excellent agreement with such a relation, as can be seen in Figure~\ref{fig:qpombh}.

One might naturally ask whether we can detect the corresponding high-frequency QPO in our data. The predicted period for the high-frequency QPO is $\sim22$~hours based on the extrapolated high-frequency relation. Although our 30-minute sampling is theoretically sensitive to such a period, it occurs in a region where the light curve is dominated by Poisson noise. In Figure~\ref{fig:qpo_powspec} we denote the location where this feature would be with an arrow. There is no detection. High-frequency QPOs can also be weaker than their low-frequency counterparts by at least a factor of ten. We therefore do not consider the lack of a 22 hour timescale to be particularly surprising. 

\section{Concluding Remarks}
\label{sec:conclusion}

Detecting a QPO in an optical light curve may have implications for some QPO models. There are two possibilities. First, the oscillations could be occurring in the optically-emitting region of the disk as well as in the inner regions assumed in X-ray studies. Second, the optical disk region may simply be reprocessing the nuclear X-ray oscillations. If the optical region of the disk is also producing oscillations, this may favor models such as density waves. In the reprocessing scenario, re-radiation from a wide range of optical disk radii may contribute to the reduced coherence of optical QPO features, as arrival times at the observer would be effectively smeared out compared with the compact X-ray region. However, the low coherence value could also be due to a wandering central frequency throughout the observation or to turbulence in the disk. Many more optical QPOs will need to be detected before they can inform interpretations of X-ray QPOs. This goal will soon be attainable if we pursue AGN science with upcoming high-cadence, long-duration timing facilities, including exoplanet-hunting satellites like TESS and PLATO.

\acknowledgments

KLS is grateful for support from the NASA Earth and Space Sciences Fellowship (NESSF). Support for this work was also provided by the National Aeronautics and Space Administration through Einstein Postdoctoral Fellowship Award Number PF7-180168, issued by the Chandra X-ray Observatory Center, which is operated by the Smithsonian Astrophysical Observatory for and on behalf of the National Aeronautics Space Administration under contract NAS8-03060. We gratefully acknowledge Tod Strohmayer for helpful discussions and the referee for remarks that significantly improved the manuscript.

\newpage

\end{document}